\documentclass[onecolumn]{article}

\usepackage{amsmath,amsfonts,showdim,graphicx,mathrsfs,subfig,psfrag,color}
\usepackage{multirow}

\newcommand{\siun}[1]{\mathrm{#1}} 

\newcommand{\CITE}[1]{\cite{#1}}
\newcommand{\re}[1]{(\ref{eq:#1})}

\def\phi{\varphi}

\def\rho{\varrho}
\def\d{\mathrm{d}}
\def\p{\partial}
\def\sgn{\mathop{\rm sgn}}
\def\sinh{\mathop{\rm sinh}}

\def\cosh{\mathop{\rm cosh}}

\def\atgh{\mathop{\rm artanh}}

\renewcommand{\vec}[1]{\boldsymbol{#1}}

\newcommand{\rem}[1]{}

\newlength{\obrA} 
\newlength{\obrB} 

\newcommand{\ssp}{\scriptscriptstyle{+}}
\newcommand{\ssm}{\scriptscriptstyle{-}}
\newcommand{\sspm}{\scriptscriptstyle{\pm}}

\newcommand{\fp}{\mathcal{A}}

\begin{document}
\bibliographystyle{spmpsci}
\title{Accelerating electromagnetic magic field from the C-metric}
\author{Ji\v{r}\'{i} Bi\v{c}\'{a}k\footnote{bicak@mbox.troja.mff.cuni.cz},\qquad David Kofro\v{n}\footnote{d.kofron@gmail.com}\vspace{1em}\\
Institute of Theoretical Physics,\\
Faculty of Mathematics and Physics, Charles University,\\
 V Hole\v{s}ovi\v{c}k\'{a}ch 2, 180\,00 Prague 8, Czech Republic
}
\maketitle
\begin{abstract}
Various aspects of the C-metric representing two rotating charged black holes accelerated in opposite directions are summarized and its limits are considered. A particular attention is paid to the special-relativistic limit in which the electromagnetic field becomes the ``magic field'' of two oppositely accelerated rotating charged relativistic discs. When the acceleration vanishes the usual electromagnetic magic field of the Kerr-Newman black hole with gravitational constant set to zero arises. 

Properties of the accelerated discs and the fields produced are studied and illustrated graphically. The charges at the rim of the accelerated discs move along spiral trajectories with the speed of light. If the magic field has some deeper connection with the field of the Dirac electron, as is sometimes conjectured because of the same gyromagnetic ratio, the ``accelerating magic field'' represents the electromagnetic field of a uniformly accelerated spinning electron. It generalizes the classical Born's solution for two uniformly accelerated monopole charges.
\end{abstract}

\emph{Keywords:} electromagnetic magic field -- Kerr-Newman solution -- C-metric -- boost-rotation symmetric spacetimes

\section{Introduction}\label{sec:Intro}
It was in the late 1950s already when J\"{u}rgen Ehlers as a member of the ``Jordan Seminar'' in the Physics Department of Hamburg University contributed substantially to the area of exact solutions of Einstein's field equations. His famous chapter \CITE{Wi} in the ``Witten book'' written together with Wolfgang Kundt became the landmark in the subject by its emphasis on characterizing exact solutions invariantly by their intrinsic geometrical properties. In a section on ``a degenerate static vacuum fields'' the authors give a table 2-3.1 of all the metrics of that type constructed originally by Levi-Civita in 1917-19. The last entry in the table consists of ``the fields of class C'' in the terminology of Ehlers and Kundt; in contrast to A and B fields the C fields do not admit an isotropy group. In the present-day terminology these vacuum solutions are called the C-metric. Ehlers and Kundt considered analytic (non-static) extensions of the fields A and B, obtaining so, e.g., the Kruskal extension of the Schwarzschild metric. 

\begin{figure}
\begin{center}
\subfloat[Space diagram]{
%
%
\begin{psfrags}%
\psfragscanon%
%
\psfrag{s05}[l][l]{\color[rgb]{0,0,0}\setlength{\tabcolsep}{0pt}\begin{tabular}{l}$z$\end{tabular}}%
\psfrag{s06}[l][l]{\color[rgb]{0,0,0}\setlength{\tabcolsep}{0pt}\begin{tabular}{l}$r$\end{tabular}}%
%
\psfrag{x01}[t][t]{-1}%
\psfrag{x02}[t][t]{-0.5}%
\psfrag{x03}[t][t]{0}%
\psfrag{x04}[t][t]{0.5}%
\psfrag{x05}[t][t]{1}%
%
\psfrag{v01}[r][r]{-1}%
\psfrag{v02}[r][r]{-0.5}%
\psfrag{v03}[r][r]{0}%
\psfrag{v04}[r][r]{0.5}%
\psfrag{v05}[r][r]{1}%
%
\includegraphics[keepaspectratio,height=6cm]{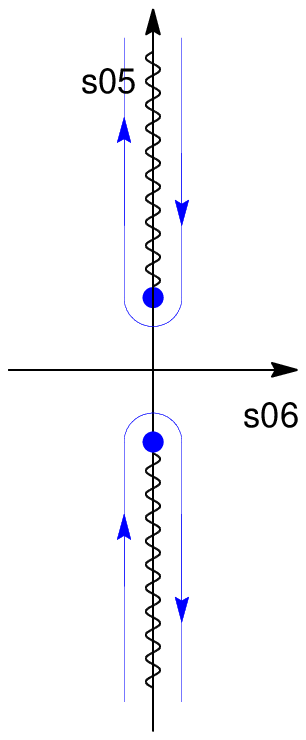}%
\end{psfrags}%
%
}\qquad\qquad
\subfloat[Spacetime diagram]{
%
%
\begin{psfrags}%
\psfragscanon%
%
\psfrag{s05}[l][l]{\color[rgb]{0,0,0}\setlength{\tabcolsep}{0pt}\begin{tabular}{l}$t$\end{tabular}}%
\psfrag{s06}[l][l]{\color[rgb]{0,0,0}\setlength{\tabcolsep}{0pt}\begin{tabular}{l}$z$\end{tabular}}%
%
\psfrag{x01}[t][t]{-1}%
\psfrag{x02}[t][t]{-0.5}%
\psfrag{x03}[t][t]{0}%
\psfrag{x04}[t][t]{0.5}%
\psfrag{x05}[t][t]{1}%
%
\psfrag{v01}[r][r]{-1}%
\psfrag{v02}[r][r]{-0.5}%
\psfrag{v03}[r][r]{0}%
\psfrag{v04}[r][r]{0.5}%
\psfrag{v05}[r][r]{1}%
%
\includegraphics[keepaspectratio,height=6cm]{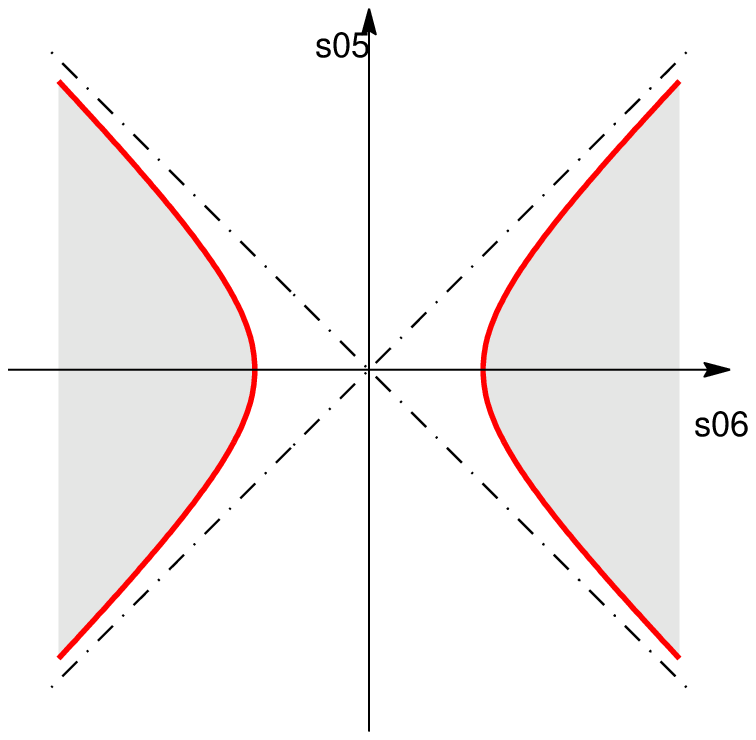}%
\end{psfrags}%
%
}
\end{center}
\caption{Space (a) and spacetime (b) diagram with two uniformly accelerated particles. The conical singularities (``strings'') are depicted in (a) by zigzag lines, their history is in the shaded region (b). The figures represent schematic diagrams also in the case of accelerated black holes. (For a detailed analysis of the conformal Penrose-Carter diagrams in the case of ``non-rotating'' C-metric, see \CITE{GKP}.)}
\label{fig:1}
\end{figure}

In 1970 Kinnersley and Walker \CITE{KW} performed the analytic extension of the C-metric. They established a clear physical interpretation of the C-metric as the Schwarzschild particles (black holes) uniformly accelerated in opposite directions (cf. Fig. \ref{fig:1}) and concluded that in some regions the C-metric has a radiative character. They also noticed similarities between the C-metric and the solutions of Bonnor and Swaminarayan \CITE{BSzp} analyzed in detail, in particular from the viewpoint of their radiative properties, by Bi\v{c}\'{a}k \CITE{Bi68} in 1968. In his work the Bondi news function for the solutions of Bonnor and Swaminarayan was derived, however, the particular forms of two functions entering the metric were not used during the derivation. The expression (26) in \CITE{Bi68} represents also the general form of the news function for the C-metric.

In fact, both the C-metric and Bonnor and Swaminarayan solutions belong to a wide class of the ``boost-rotation symmetric spacetimes'' representing the exterior fields of uniformly accelerated sources\footnote{Notice, however, that until now no interior exact solution was found except for black holes -- the case of the C-metric. Recently, Bernd Schmidt has been working on the proof of the existence of such interior solutions.} in general relativity. The encouragement and kind support of J\"{u}rgen Ehlers were important factors in our collaboration with Bernd Schmidt which led to the paper \CITE{BiSchm} dealing for the first time with boost-rotation symmetric spacetimes from a unified point of view. These spacetimes are the only explicitly known solutions of the vacuum Einstein field equations which describe moving finite objects, are radiative and asymptotically flat in the sense that they admit global, though not complete, smooth null infinity, as well as spacelike and timelike infinities. For a brief survey of the main properties of these solutions see, for example, the chapter ``Selected Solutions of Einstein's Field Equations: Their Role in General Relativity and Astrophysics'' \CITE{BiE} in the volume dedicated to J\"{u}rgen Ehlers on the occasion of his 70th birthday.

A number of papers were written on various specific aspects of the boost-rotation symmetric solutions, in particular on the C-metric. In our very recent work \CITE{BiKo} dedicated to the memory of J\"{u}rgen Ehlers we analyzed the generic spacetimes with axial and boost symmetries from the Newtonian perspective by employing the Ehlers frame theory \CITE{Ehl-ex}, \CITE{EhlN}, to construct the Newtonian limit rigorously. This work corroborated the physical significance of the boost-rotation symmetric spacetimes by demonstrating that the \emph{Newtonian limit} corresponds to the gravitational field of classical point masses accelerated uniformly in the classical mechanics. We illustrated the results by examples such as the C-metric.

In the present paper we shall, among others, discuss the \emph{special-relativistic} limit of a specific boost-rotation symmetric solution -- of the charged rotating C-metric representing two uniformly accelerated, charged and rotating (Kerr-Newman) black holes.

In most of the work on the generic boost-rotation symmetric spacetimes one assumes the boost Killing vector $\vec{\xi} = z\vec{\p_t}+t\vec{\p_z}$ (considering the boosts along the $z$-axis) and the axial Killing vector $\vec{\eta}=\vec{\p_\phi}$ to be hypersurface orthogonal. The general boost-rotation symmetric spacetimes are of Petrov type I and it appears too difficult to analytically find their extension to the cases when $\vec{\p_\phi}$ is not hypersurface orthogonal or an electromagnetic field is present since in such cases none of the Einstein field equations reduces to a linear equation like in the hypersurface orthogonal cases. However, in the case of the C-metric which is algebraically degenerate (of Petrov type D) the rotating and charged solutions were found to be contained in a general class of type D solutions to the Einstein-Maxwell equations discovered in 1976 by Pleba\'{n}ski and Demia\'{n}ski \CITE{PD}.

Their class contains seven real parameters -- two parameters describe the mass and the NUT parameter, two parameters are related to the angular momentum per unit mass and acceleration, two parameters to the electric and magnetic charges and the last is the cosmological constant \CITE{PD}, \CITE{SC}. Setting the cosmological constant, magnetic charge and NUT parameter to zero, we should get a charged, rotating and accelerating object. 

However, for a long time both the vacuum C-metric and its generalizations were  analyzed in coordinate systems unsuitable for treating global issues since these systems were adapted to the degenerate character of the metric. A transformation which brings the rotating vacuum C-metric into the canonical form of radiative spacetimes with boost-rotation symmetry (geometrically introduced first in the nonrotating case in \CITE{BiSchm}) was found in \CITE{BiPr}. By analytically continuing the metric across ``acceleration horizons'', two new regions of spacetime arise in which both Killing vectors are spacelike and the metric can be shown to represent two uniformly accelerated, rotating black holes, either connected by a conical (nodal) singularity, or with conical singularities extending from each of them to infinity. By plotting curvature invariants one can show how the gravitational pulse propagates in all directions from black holes (see Fig. 5 in \CITE{BiPr} where also references to other related papers are given).

It was also noticed in \CITE{BiPr} that there exists causality violation region in the neighbourhood of the nodal singularity which tends to be dragged along with rotating black holes. There are in general so-called torsion singularities \CITE{LeOl}, \CITE{Bo2003} associated with the conical singularities in the usual form of the rotating C-metric considered in \CITE{BiPr}. In 2003 Hong and Teo \CITE{HT1} proposed a new form of the standard (non-rotating) C-metric in which the ``structure function'' determining the metric was explicitly factorizable; their new form is related to the usual one by a mere coordinate transformation. In a subsequent paper \CITE{HT2} Hong and Teo introduced an analogous new form of the rotating C-metric. However, in contrast to the non-rotating case, this is physically distinct from the usual form: the conical singularities are free of torsion singularities and the causality is not violated. Technically, their ``new form'' is again a member of the Pleba\'{n}ski-Demia\'{n}ski class, the ``usual'' and the new form differ in the choice of the parameter traditionally interpreted as the ``NUT parameter''. Before their work such interpretation was accepted on the basis of a non-trivial limiting procedure bringing the C-metric to the spacetime of an unaccelerated source. For further elaborations on the Hong's and Teo's work see \CITE{GrPo}.

It the present work in Section \ref{sec:rCM}  we give the metric representing two accelerating, rotating and charged black holes in the new form of Hong and Teo and consider its various limits. We explicitly demonstrate how (i) the Kerr-Newman solution arises when the acceleration parameter is set to zero, (ii) with zero rotation the standard C-metric follows, (iii) when the gravitational constant is send to zero, the ``accelerated electromagnetic magic field'' results, (iv) if, in addition the acceleration parameter is set to zero, the ``usual'' electromagnetic magic field representing the charged uniformly rotating conducting stationary disc arises, and (v) in the limit of vanishing both gravitational constant and the rotation parameter we get the well known classical Born's solution for the point monopole charges of equal magnitude but opposite signs uniformly accelerated in opposite directions along timelike orbits of the boost Killing vector field in flat spacetime (Fig. \ref{fig:1}).

What is the ``electromagnetic magic field''? The term was coined by Lynden-Bell \CITE{DLB-Magic} in his thorough analysis of a closed form analytic solutions of Maxwell's equations for the relativistically uniformly rotating conducting charged disc for all values of the tip speed up to the velocity of light. When velocity at the rim reaches $v=c$ the field becomes the magic electromagnetic field of the Kerr-Newman solution with gravitational constant $G$ set to zero. The field can be obtained by considering the potential of a point charge located at a complex position $\vec{r}_0+i\vec{a}$ in flat space. Choosing $\vec{r}_0$ at the origin and $\vec{a}$ along the $z$-axis we get $\psi = q/(r^2-2iar\cos\theta-a^2)^{1/2}$ and the magic field is given simply by $\vec{F}=  \vec{E}+i\vec{B} = -\nabla\psi$.

Let us summarize some basic properties of the magic field as described in \CITE{DLB-Magic}, \CITE{DLB-ElM-PrD}. The field has total charge $q$, magnetic dipole $qa$ and non-vanishing only even electric and odd magnetic multipole moments. The sources of the field lie on the singular ring at $\rho=a$, $z=0$ and on the disc $\rho<a$, $z=0$ (in cylindrical coordinates). The electric field is orthogonal to the disc, so the disc is conducting. The surface density of charge on the disc is of the opposite sign to the total charge. The singular ring has the (infinite) charge opposite to that in the disc such that the total charge equals $q$. The charge density is rotating rigidly with the angular velocity $\omega=c/a$ but the current moving with velocity $v=c$ around the singular ring is of opposite sign. Lynden-Bell \CITE{DLB-Magic} (see also \CITE{DLB-ElM-PrD}) lists a number of other properties of the magic field like, for example, its invariants, the field energy density, the Poynting vector and others. In a subsequent paper \CITE{LB2} Lynden-Bell constructed the field of a relativistically spinning charged sphere and showed that when the equatorial velocity approaches $c$ the charge of the same sign as the total charge concentrates in an equatorial belt whereas the charge of opposite sign lies on the most of the sphere. The structure of the field starts to resemble the magic field.

Among main remarkable properties of the magic field is the fact that it has gyromagnetic ratio the same as the Dirac electron and that wave equations and Dirac equation are separable in this field. There is no place here to summarize the literature on various aspects of these properties. As an illustration let us just quote more recent work by Newman \CITE{New} on the classical geometrical origin of the Dirac gyromagnetic ratio and the last of the series of papers by Pekeris and Frankowski \CITE{PeFr} in which the atomic nucleus is represented as a Kerr-Newman source and the hyperfine splitting in muonium, positronium and hydrogen is studied. 

\begin{figure}
\begin{center}
%
%
\begin{psfrags}%
\psfragscanon%
%
\psfrag{s05}[l][l]{\color[rgb]{0,0,0}\setlength{\tabcolsep}{0pt}\begin{tabular}{l}{\scriptsize \emph{Z}}\end{tabular}}%
\psfrag{s06}[l][l]{\color[rgb]{0,0,0}\setlength{\tabcolsep}{0pt}\begin{tabular}{l}{\scriptsize \emph{X}}\end{tabular}}%
%
\psfrag{x01}[t][t]{{\scriptsize -2}}%
\psfrag{x02}[t][t]{{\scriptsize -1}}%
\psfrag{x03}[t][t]{{\scriptsize 0}}%
\psfrag{x04}[t][t]{{\scriptsize 1}}%
\psfrag{x05}[t][t]{{\scriptsize 2}}%
%
\psfrag{v01}[r][r]{{\scriptsize -2}}%
\psfrag{v02}[r][r]{{\scriptsize -1}}%
\psfrag{v03}[r][r]{{\scriptsize 0}}%
\psfrag{v04}[r][r]{{\scriptsize 1}}%
\psfrag{v05}[r][r]{{\scriptsize 2}}%
%
\includegraphics[keepaspectratio,width=.45\textwidth]{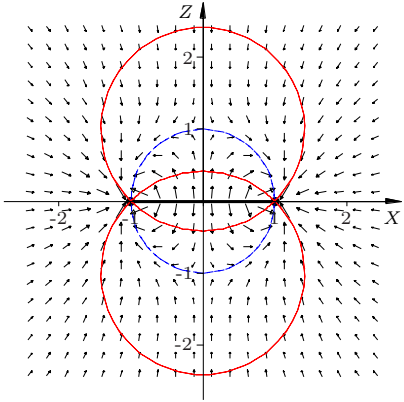}%
\end{psfrags}%
%

\end{center}
\caption{The magic electric field produced by a relativistically rotating charged disc (indicated by a thick line segment) at rest in the equatorial plane. The rim of the disc is formed by charges of opposite sign than those on the disc (notice the directions of ``arrows''). The charges at the rim move with the speed of light. By thinner, respectively dashed lines are indicated the places where the invariants $\vec{E}^2-\vec{B}^2$, resprespectively $\vec{E}\cdot\vec{B}$, vanish. (See Introduction for more details.)  }
\label{fig:MF}
\end{figure}

In the following section we first give the accelerating and rotating charged C-metric and the corresponding electromagnetic potential in the form presented by Hong and Teo \CITE{HT2}. After a coordinate transformation which enables us to perform special-relativistic limit in a simple way we calculate the electromagnetic field tensor and its invariants. Various limiting cases of the C-metric and electromagnetic field are then studied systematically. The details on the limits yielding the Kerr-Newman metric and the ``standard'' magic field are contained in Appendix \ref{app:limA}; how the classical Born's solution for two uniformly accelerated charges arises is demonstrated in Appendix \ref{app:Born}.

In Section \ref{sec:AccMF} the accelerating magic field is presented and its properties are discussed both analytically and graphically. It corresponds to two rotating charged conducting discs uniformly accelerated in opposite directions. The discs are causally disconnected like the point charges in the Born's solution. The discs are bent in the direction opposite to their acceleration. The outer rings move with higher accelerations than the central parts.

In the last Section \ref{sec:sources} the sources of the accelerating magic field are studied in more detail. We find, for example, that the charges at the rim of the discs move along the spiral trajectories with the speed of light. Their sign is opposite to that of the surface charges on the disc. These features are similar to those of the stationary magic field.

Various aspects of the accelerating magic fields remain to be explored. For example, in the region above the acceleration horizons (``above the roof'' in the terminology of \CITE{BiSchm}) where the boost Killing vector is spacelike, the field is fully dynamical. Radiative properties of the accelerated magic field are of particular interest. In order to explore them the form of the fields of uniformly accelerated multipoles \CITE{BM} should be useful.

\section{Uniformly accelerated Kerr-Newman black holes} \label{sec:rCM}
As discussed in the Introduction, the most natural generalization of the standard C-metric describing two ``uniformly accelerated Schwarzschild black holes'' in opposite directions to the case when the black holes are charged and rotating appears to be given by Hong and Teo \CITE{HT2}. The charged rotating C-metric in their description reads (cf. Eqs. (22), (23) and (15) in \CITE{HT2}):
\begin{multline}
\d s^2 = \frac{1}{A^2(x-y)^2}\;\biggl\{ \frac{\mathcal{G}(y)}{1+\left( aAxy \right)^2}\Bigl[ \d t+aA\left( 1-x^2 \right)K\d\phi \Bigr]^2-\frac{1+\left( aAxy \right)^2}{\mathcal{G}(y)}\,\d y^2 \\
+\frac{1+\left( aAxy \right)^2}{\mathcal{G}(x)}\,\d x^2 +\frac{\mathcal{G}(x)}{1+\left( aAxy \right)^2}\;\Bigl[ \left( 1+a^2A^2y^2 \right)K\d\phi+aAy^2\d t \Bigr]^2\biggr\}\,,
\label{eq:rCM}
\end{multline}
where the structure function $\mathcal{G}$ is defined by
\begin{equation}
\mathcal{G}(\xi) = \left( 1-\xi^2 \right)\left( 1+r_{\ssp}A\xi \right)\left( 1+r_{\ssm}A\xi \right)
\label{eq:rG}
\end{equation}
with
\begin{equation}
r_\pm=G m \pm \sqrt{G^2m^2-a^2-G q^2}\,.
\label{eq:rrpm}
\end{equation}
Here $m$, $a$, $q$ and $A$ are respectively the mass, rotation, charge and acceleration parameter. In addition to the form given in \CITE{HT2} we introduced constant $K$ scaling the angular coordinate $\phi$ and inserted explicitly the Newtonian gravitational constant $G$ (we put the speed of light $c=1$, except for Appendix \ref{app:Born}). The possibility to cast the structure function $\mathcal{G}$ to explicitly factorizable form \re{rG} was the main achievement in \CITE{HT2}. It enables one to find easily four simple real roots of $\mathcal{G}$,
\begin{equation}
\xi_1 = -\frac{1}{r_{\ssm}A}\,,\quad \xi_2 = -\frac{1}{r_{\ssp}A}\,,\quad \xi_3 = -1\,,\quad \xi_4 = 1\,.
\label{eq:realroots}
\end{equation}
(Recall that in the original Demia\'{n}ski-Pleba\'{n}ski form \CITE{PD} this structure function is the quartic polynomial with very complicated roots.) The roots obey $\xi_1\leq\xi_2<\xi_3<\xi_4$ and they determine the allowed range of $x$ and $y$ coordinates in \re{rG} as follows: $\xi_2\leq y \leq \xi_3$ and $\xi_3\leq x \leq \xi_4$. The lines $x=\xi_4$ and $x=\xi_3$ are parts of the symmetry axis, the black hole horizon is at $y=\xi_2$, the acceleration horizon ($t=\pm z$ in Fig. \ref{fig:1}) is at $y=\xi_3$ (see \CITE{HT2} for more details).

The electromagnetic 4-potential is given by (cf. Eq. (12) in \CITE{HT2})
\begin{equation}
\vec{\fp} = \frac{qy}{1+\left( aAxy \right)^2}\,\left[ \vec{\d} t+aA\left( 1-x^2 \right)K\vec{\d}\phi \right]\,.
\label{eq:A}
\end{equation}

The axial Killing vector field is simply $\vec{\eta}=\vec{\p_\phi}$, with the norm $F=\eta^a\eta_a$. The axis of symmetry is regular if in the limit at the rotation axis $F^{,a}F_{,a}/4F \rightarrow 1$ (see, e.g., \CITE{SC}, Eq. (19.3)). Calculating this invariant we get
\begin{equation}
\frac{F^{,a}F_{,a}}{F} = \frac{A^2\left( x-y \right)^2}{1+\left( aAxy \right)^2} \frac{\bigl( \mathcal{G}(x)F^2_{,x}-\mathcal{G}(y)F^2_{,y} \bigr)}{F}\,.
\label{eq:}
\end{equation}

Since parts of the axis of symmetry are given by the roots of the structure function $x=\xi_3$ and $x=\xi_4$, and these are $\xi_3=-1$ and $\xi_4=1$ (see \re{realroots}), the condition of the axis regularity turns out to be
\begin{equation}
\frac{1}{4}\frac{F^{,a}F_{,a}}{F}\quad \stackrel{x\rightarrow \pm 1}{\longrightarrow}\quad K^2 \left( 1+a^2A^2 +G\left( A^2q^2\pm 2Am \right)\right)^2 \quad\longrightarrow\quad 1\,.
\label{eq:rDefAo} 
\end{equation}
Obviously by choosing a free constant $K$ appropriately we can make the axis regular either between the black holes (at $x=1$) or from each of the black holes to infinity ($x=-1$). If we are primarily interested just in the special relativistic limit, $G\rightarrow 0$, which results in the Minkowski space (thought not necessarily in the Lorentzian coordinates -- see below) without any conical singularity, we can put
\begin{equation}
K = \left( 1+a^2A^2 \right)^{-1}\,.
\label{eq:Kl}
\end{equation}
However, we may start from the full general-relativistic metric, make one of the part of the axis regular by choosing either $K=K_{\ssp}$ or $K=K_{\ssm}$ with
\begin{equation}
K_{\sspm} = \left[ 1+a^2A^2+G\left( A^2q^2\pm 2Am \right) \right]^{-1}\,,
\label{eq:Kf}
\end{equation}
and going then over to the special-relativistic limit $G\rightarrow 0$ we still get smooth Minkowski spacetime without conical singularities.

It is still useful to make a simple coordinate transformation in the metric \re{rCM} corresponding to a rigid rotation
\begin{equation}
\phi' = \phi -\omega_0 t\,,\qquad t' = t_0 t\,,
\label{eq:cs}
\end{equation}
where constants $\omega_0$ and $t_0$ are given by
\begin{equation}
\omega_0 = \frac{aA}{K\left( 1+a^2A^2 \right)}\,,\qquad t_0 = K\left( 1+a^2A^2 \right)\,.
\label{eq:}
\end{equation}
This transformation removes the non-diagonal term $g_{t\phi}$ which would other\-wise appear in the flat-space limit $G\rightarrow 0$ of the metric \re{rCM}. Omitting the primes the transformation \re{cs} brings the full accelerating, rotating and charged C-metric into the form
\begin{multline}
\d s^2 = \frac{1}{A^2(x-y)^2}\,\biggl\{ \frac{\mathcal{G}(y)}{1+\left( aAxy \right)^2}\;\Bigl[ \left( 1+a^2A^2x^2 \right)K\d t+aA\left( 1-x^2 \right)K\d\phi \Bigr]^2 \\
-\frac{1+\left( aAxy \right)^2}{\mathcal{G}(y)}\,\d y^2 +\frac{1+\left( aAxy \right)^2}{\mathcal{G}(x)}\,\d x^2\\
+\frac{\mathcal{G}(x)}{1+\left( aAxy \right)^2}\;\Bigl[ \left( 1+a^2A^2y^2 \right)K\d\phi+aA\left(y^2-1\right)K\d t \Bigr]^2\biggr\}
\label{eq:rCMr}
\end{multline}
with the electromagnetic 4-potential
\begin{equation}
\vec{\fp} = \frac{Kqy}{1+\left( aAxy \right)^2}\,\left[ \left(1+a^2A^2x^2\right)\vec{\d} t+aA\left( 1-x^2 \right)\vec{\d}\phi \right]\,.
\label{eq:rA}
\end{equation}
The electromagnetic field reads as follows:
\begin{multline}
\vec{F} = \frac{Kq}{\left[1+\left( aAxy \right)^2\right]^2}\Biggl[ 2A^2a^2xy\left( y^2-1 \right)\vec{\d}t\wedge\vec{\d}x \\
+ \left( 1+\left( aAx \right)^2 \right)\left( -1+\left( aAxy \right)^2 \right)\vec{\d}t\wedge\vec{\d}y \\
- 2aAxy\left( 1+\left( aAy \right)^2 \right)\vec{\d}x\wedge\vec{\d}\phi + aA\left( 1-x^2 \right)\left( 1-\left( aAxy \right)^2 \right)\vec{\d}y\wedge\vec{\d}\phi\Bigg]\,.
\label{eq:elm}
\end{multline}

The electromagnetic invariants are found to be
\begin{eqnarray}
\frac{1}{2}\,F_{ab}F^{ab} &=& -\frac{q^2A^4\left( x-y \right)^4\left[ 1-6\left( aAxy \right)^2+\left( aAxy \right)^4 \right]}{\left[1+\left( aAxy \right)^2 \right]^4}\,, \label{eq:inv1}\\
\frac{1}{4}\,F_{ab} \star F^{ab} &=& -\frac{2q^2A^4\left( x-y \right)^4aAxy\left[ -1+\left( aAxy \right)^2 \right]}{\left[ 1+\left( aAxy \right)^2 \right]^4}\,.
\label{eq:inv2}
\end{eqnarray}

Now starting from the metric \re{rCMr} and electromagnetic potential \re{rA} describing the charged, rotating and accelerating black holes we can go over to simpler, more familiar cases by making appropriate limits:
\begin{enumerate}
\item Putting the rotational parameter $a=0$, the expressions \re{rCMr}, \re{rA} become directly the metric and potential of two accelerated non-rotating charged black holes (cf. Eqs. (6a)-(8) in \CITE{HT1}).
\item Taking $a=0$ and $G\rightarrow 0$, we get the Minkowski metric and the field of two opposite charges uniformly accelerated in opposite directions -- the classical Born's solution, as described in Appendix \ref{app:Born}.
\item From the Born solution one can obtain the uniform electric field by increasing the magnitude of the opposite charges and, simultaneously, their distance. Recently, more sophisticated limit was performed to construct the Melvin (electrical) universe, i.e., gravitating ``uniform'' electric field, from the charged, non-rotating C-metric \CITE{HaKr}.
\item In the limit of vanishing acceleration, $A\rightarrow 0$, the Kerr-Newman solution in the usual Boyer-Lindquist coordinates (see, e.g., \CITE{MTW}) is recovered after a suitable coordinate transformation (see Appendix \ref{app:limA}).
\item Putting $G=0$ in the zero acceleration limit $A\rightarrow 0$, we get the electromagnetic magic field (see, e.g., \CITE{DLB-ElM-PrD}) of the Kerr-Newman solution.
\item Putting $G\rightarrow 0$ but leaving $A\neq 0$, $a\neq 0$ and $q\neq 0$, we find the ``accelerating electromagnetic magic field'' which has not been considered in the literature so far. In the following we shall study its properties in some detail.
\end{enumerate}

\section{Accelerated magic field} \label{sec:AccMF}
Taking the limit $G\rightarrow 0$ in the metric \re{rCMr} while keeping $a$, $A$, $q$ constant and choosing constant $K$ as in \re{Kl}, we arrive at flat spacetime in non-trivial coordinates: 
\begin{multline}
\d s^2 = \frac{1}{A^2\left( x-y \right)^2}\biggl[ -\frac{\left( y^2-1 \right)\left( 1+a^2A^2x^2 \right)}{1+a^2A^2}\,\d t^2 + \frac{1+\left( aAxy \right)^2}{\left( 1-x^2 \right)\left( 1+a^2A^2x^2 \right)}\,\d x^2 \\
+\frac{1+\left( aAxy \right)^2}{\left( y^2-1 \right)\left( 1+a^2A^2y^2 \right)}\,\d y^2 + \frac{\left( 1-x^2 \right)\left( 1+a^2A^2y^2 \right)}{1+a^2A^2}\,\d\phi^2\biggr]\,.
\label{eq:accOSph}
\end{multline}
The form of the electromagnetic potential, the field and its invariants remain the same as in Eq. \re{rA} and \re{elm}.

In the flat spacetime limit the coordinates ${t,\,x,\,y,\,\phi}$ turn out to be just ``complicated'' coordinates in a uniformly accelerated frame. In such a frame the standard convention is to use the Rindler coordinates (see, e.g., \CITE{MTW}) in which the metric reads 
\begin{equation}
\d s^2 = -\zeta^2\d t^2 + \d\zeta^2+\d\rho^2+\rho^2\d\phi^2\,.
\label{eq:rFlat}
\end{equation}
By making the coordinate transformation
\begin{align}
\zeta &= \frac{\sqrt{\left( y^2-1 \right)\left( a^2A^2x^2+1 \right)}}{A\left( x-y \right)\Gamma}\,, & 
\rho &= \frac{\sqrt{\left( 1-x^2 \right)\left( a^2A^2y^2+1 \right)}}{A\left( x-y \right)\Gamma}\,, \\
t &=t\,, & \phi&=\phi\,,
\label{eq:rTrf}
\end{align}
where we denoted $\Gamma = \sqrt{1+a^2A^2}$, we obtain directly the metric in the form \re{accOSph}.

The inverse transformation is more complicated (but it is useful to know and it is necessary to draw numerically the figures below):
\begin{multline}
x = \frac{-1}{2} 
\frac{1}
{\sgn (\xi_{\ssm})\ aA\xi_{\ssp}} \Biggl[ -4\left( \xi_{\ssm}^2+4\rho^2/A^2 \right) \\
+\left( \sqrt{\Gamma^2\xi_{\ssp}^2-4\left( \zeta/A+a\rho \right)^2 }+\sqrt{\Gamma^2\xi_{\ssp}^2-4\left( \zeta/A-a\rho \right)^2} \right)^2\Biggr]^{1/2}\,, \label{eq:xinv}
\end{multline}
\begin{multline}
y = \frac{-1}{2} \frac{1} {\sgn (\xi_{\ssm})\ aA\xi_{\ssm}} 
\Biggl[ -4\left( \xi_{\ssm}^2+4\rho^2/A^2 \right) \\
+\left( \sqrt{\Gamma^2\xi_{\ssm}^2+4\left( \rho/A-a\zeta\right)^2 }+\sqrt{\Gamma^2\xi_{\ssm}^2+4\left( \rho/A+a\zeta\right)^2} \right)^2 \Biggr]^{1/2}\,, \label{eq:yinv}
\end{multline}
where
\begin{equation}
\xi_{\sspm} = \rho^2+\zeta^2\pm\frac{1}{A^2} \,.
\label{}
\end{equation}

\begin{figure}
\begin{center}
%
%
\begin{psfrags}%
\psfragscanon%
%
\psfrag{s05}[l][l]{\color[rgb]{0,0,0}\setlength{\tabcolsep}{0pt}\begin{tabular}{l}{\scriptsize \emph{Z}}\end{tabular}}%
\psfrag{s06}[l][l]{\color[rgb]{0,0,0}\setlength{\tabcolsep}{0pt}\begin{tabular}{l}{\scriptsize \emph{X}}\end{tabular}}%
%
\psfrag{x01}[t][t]{{\scriptsize -0.8}}%
\psfrag{x02}[t][t]{{\scriptsize  -0.6}}%
\psfrag{x03}[t][t]{{\scriptsize  -0.4}}%
\psfrag{x04}[t][t]{{\scriptsize -0.2}}%
\psfrag{x05}[t][t]{{\scriptsize 0}}%
\psfrag{x06}[t][t]{{\scriptsize 0.2}}%
\psfrag{x07}[t][t]{{\scriptsize 0.4}}%
\psfrag{x08}[t][t]{{\scriptsize 0.6}}%
\psfrag{x09}[t][t]{{\scriptsize 0.8}}%
%
\psfrag{v01}[t][t]{{\scriptsize -0.8}}%
\psfrag{v02}[t][t]{{\scriptsize -0.6}}%
\psfrag{v03}[t][t]{{\scriptsize -0.4}}%
\psfrag{v04}[t][t]{{\scriptsize -0.2}}%
\psfrag{v05}[t][t]{{\scriptsize 0}}%
\psfrag{v06}[t][t]{{\scriptsize 0.2}}%
\psfrag{v07}[t][t]{{\scriptsize 0.4}}%
\psfrag{v08}[t][t]{{\scriptsize 0.6}}%
\psfrag{v09}[t][t]{{\scriptsize 0.8}}%
%
\includegraphics[keepaspectratio,width=0.9\textwidth]{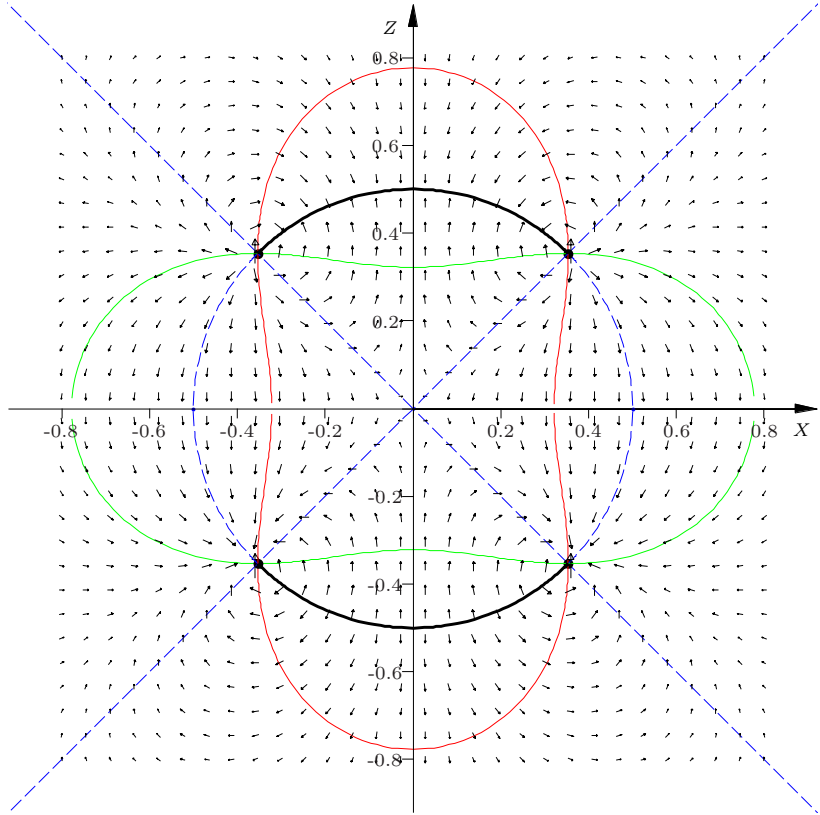}%
\end{psfrags}%

\end{center}
\caption{The electric intensity of the ``accelerated magic field'' produced by two relativistically rotating discs uniformly accelerated in opposite directions. The discs, indicated by thick lines, are bent in the direction opposite to the acceleration in the frames, in which they are momentarily at rest. At large times when all their parts approache the velocity of light in a given inertial frame, the discs become flattened. By thinner, respectively dashed lines are indicated the places where the invariants $\vec{E}^2-\vec{B}^2$, respectively $\vec{E}\cdot\vec{B}$, vanish. The discs and the electric field are here plotted at time $T=0$ when the discs are momentarily at rest. (For more details on the properties of the field, see the text.)}
\label{fig:accMF0}
\end{figure}

\begin{figure}
\begin{center}
\subfloat[]{\label{3b}
%
%
\begin{psfrags}%
\psfragscanon%
%
\psfrag{s05}[l][l]{\color[rgb]{0,0,0}\setlength{\tabcolsep}{0pt}\begin{tabular}{l}{\scriptsize \emph{Z}}\end{tabular}}%
\psfrag{s06}[l][l]{\color[rgb]{0,0,0}\setlength{\tabcolsep}{0pt}\begin{tabular}{l}{\scriptsize \emph{$\rho$}}\end{tabular}}%
%
\psfrag{x05}[t][t]{{\scriptsize 0}}%
\psfrag{x06}[t][t]{{\scriptsize 0.2}}%
\psfrag{x07}[t][t]{{\scriptsize 0.4}}%
\psfrag{x08}[t][t]{{\scriptsize 0.6}}%
\psfrag{x09}[t][t]{{\scriptsize 0.8}}%
%
\psfrag{v06}[t][t]{{\scriptsize 0.2}}%
\psfrag{v07}[t][t]{{\scriptsize 0.4}}%
\psfrag{v08}[t][t]{{\scriptsize 0.6}}%
\psfrag{v09}[t][t]{{\scriptsize 0.8}}%
%
\includegraphics[keepaspectratio,height=0.45\textwidth]{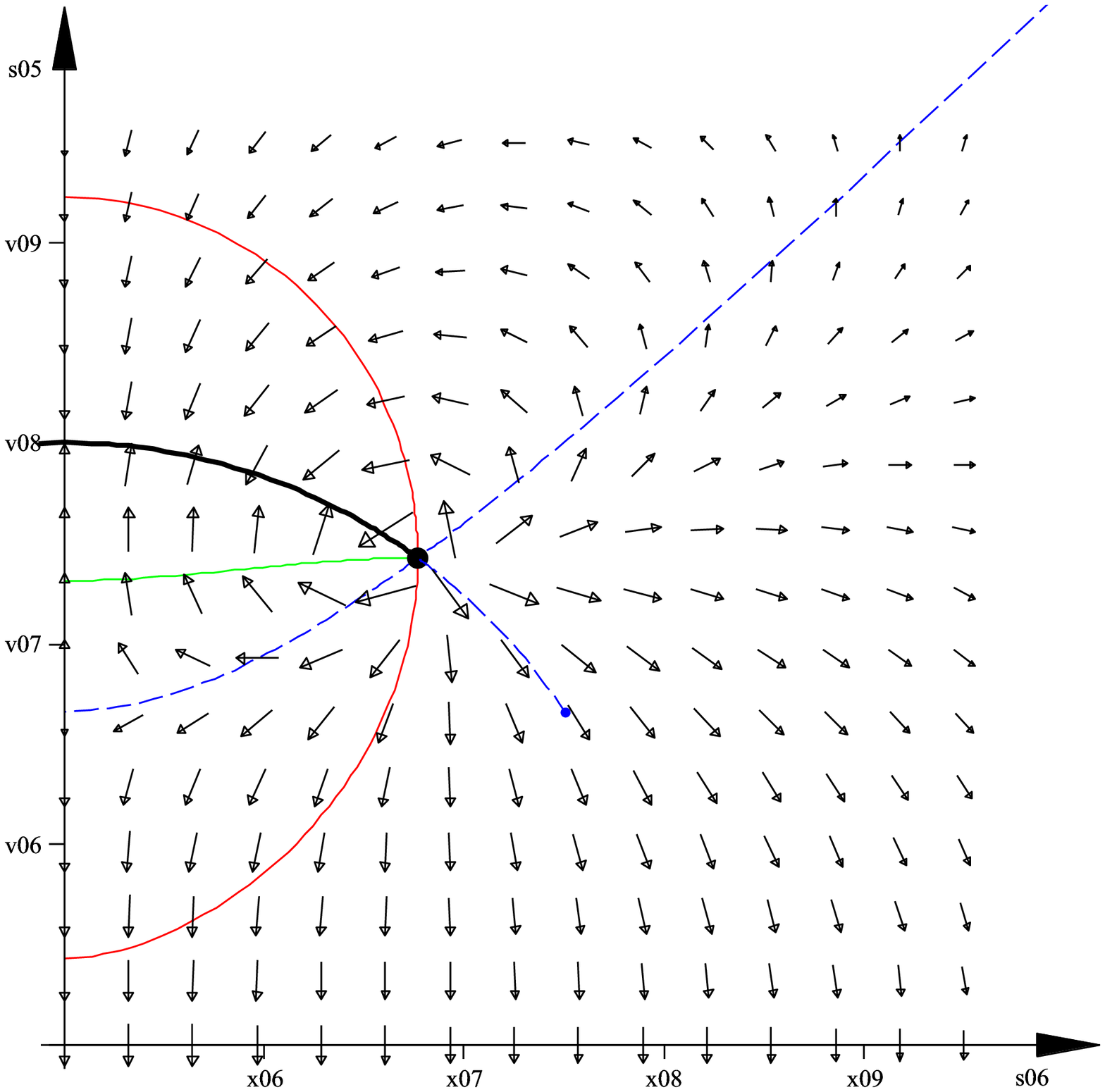}%
\end{psfrags}%
}
\subfloat[]{\label{spiral}\includegraphics[height=0.45\textwidth, keepaspectratio]{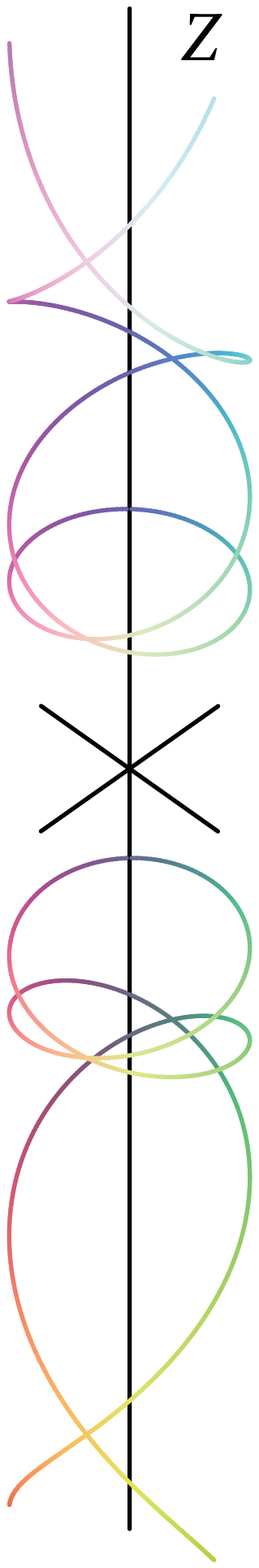}}
\end{center}
\caption{(a) The disc and the accelerating electric magic field at time $T=1/3$  -- only one quadrant of Fig. \ref{fig:accMF0} is plotted. See the caption of Fig. \ref{fig:accMF0} for more details. (b) Individual charged particles of the disc move along spiral trajectories. The particles at the rim move with the speed of light.}
\label{fig:accMF}
\end{figure}

The source of the field itself is located at $y=-\infty$. This means that its worldtube is given by
\begin{eqnarray}
\zeta &=& \frac{\sqrt{1+a^2A^2x^2}}{A\Gamma}\,, \label{eq:discZ}\\
\rho &=& \frac{a\sqrt{1-x^2}}{\Gamma}\,, \label{eq:discR}
\label{eq:}
\end{eqnarray}
where $x\in\langle -1,\,1\rangle$ and $t\in \mathbb{R}$ and $\phi\in (0\,,2\pi\rangle$. In Minkowski coordinates we have $Z=\sqrt{\zeta^2+T^2}$.
Expressing $\zeta$ from \re{discZ}\,--\,\re{discR} directly as a function of $\rho$, we find the disc to be described by the formula
\begin{equation}
\zeta = \frac{\sqrt{1-A^2\rho^2}}{A}\,.
\label{eq:fromG}
\end{equation}
Thus the disc is convexly bent ``oppositely'' to the direction of the acceleration (see Figs. \ref{fig:accMF0}, \ref{fig:accMF}\subref{3b}). 

Since the disc is bent, its different rings are moving with different accelerations. Hence, we parametrize worldlines of particular particles of the disc as follows 
\begin{eqnarray}
T &=& \tilde{B}^{-1}(\rho)\sinh \tilde{A}(\rho)\tau\,, \\
Z &=& \tilde{B}^{-1}(\rho)\cosh \tilde{A}(\rho)\tau\,, \\ 
\phi &=& \phi(\tau,\,\rho)\,, 
\label{}
\end{eqnarray}
where $0\leq\rho\leq a\Gamma^{-1}$ and $\tau$ is the proper time.
For the disc described by Eq. \re{fromG} we get $\zeta^2=Z^2-T^2=\tilde{B}^{-2}=A^{-2}\left( 1-A^2\rho^2 \right)$.

The 4-velocity of each element of the disc must satisfy the normalisation condition ($\,\dot{}=\d/\d\tau$)
\begin{equation}
-\dot{T}^2+\dot{Z}^2+R^2\dot{\phi}^2 = -\left( \tilde{A}\tilde{B}^{-1} \right)^2 + \rho^2\dot{\phi}^2 = -1\,,
\label{eq:rNC}
\end{equation}
so that
\begin{equation}
\tilde{A}^2(\rho) = A^2\left( \frac{1+\rho^2\dot{\phi}^2(\rho)}{1-A^2\rho^2} \right).
\label{eq:Arho}
\end{equation}
To determine the dependence $\dot{\phi}(\tau,\rho)$ we need to take the electromagnetic field produced by the disc into account. This will be done in the following section.

The accelerating electromagnetic magic field \re{elm} is determined by the potential \re{rA} and by the field tensor \re{elm}. Although it can be given in an analytic form in the Rindler coordinates using the transformation \re{xinv}\,--\,\re{yinv}, its form is very complicated and we will not write it explicitly down here. Its character is well seen in Figs. \ref{fig:accMF0} and \ref{fig:accMF}\subref{3b} where the accelerated magic field is plotted in different times by using Maple. It should be compared with the stationary magic field of the Kerr-Newman source plotted in Fig. \ref{fig:MF}.

In particular, notice from Fig. \ref{fig:accMF0} that we get two discs accelerating in opposite directions along the $Z$-axis. As with Born's solution for accelerating point charges of opposite signs (see Appendix \ref{app:Born} and, e.g., \CITE{R}, \CITE{Bi68}). The distribution of the charges on the discs is opposite. The disc moving along $Z>0$ is positively charged at the rim (the field points away from the rim), whereas the field lines point towards the inner part of the disc perpendicularly, indicating a negative charge surface density distribution on the disc. Inspecting the direction of the electric field in the neighbourhood of the disc accelerated along $Z<0$ we find the situation exactly opposite. This behaviour corresponds to the fact that the field is analytic everywhere except for the places where the sources (the discs) occur. Just one half ot the ``upper disc'' and the surrounding magic field is plotted in \ref{fig:accMF}\subref{3b}. As the discs move towards the null infinity their shape in a given inertial frame gets flattened. 

\rem{
\begin{equation}
F_{\alpha\beta} = \begin{pmatrix} 
0 & -E_R & -E_Z & -E_\phi \\
E_R & 0 & B_\phi /R & -RB_Z \\
E_Z & -B_\phi /R & 0 & RB_R \\
E_\phi & RB_Z & -RB_R & 0
\end{pmatrix}
\label{eq:Fcyl}
\end{equation}
This may seem surprisingly, but it is quite natural, if we consider that we have accelerating source of electric and \emph{magnetic} field. Magnetic field has its radial component, and this generates axial electric field in uniformly moving coordinate system (as can be seen from Lorentz transformation: $\vec{F} = \dots +E_\phi\vec{\d T}\wedge \vec{\d\phi}+RB_R\vec{\d Z}\wedge \vec{\d\phi} \ \rightarrow \ \dots +(E_\phi-\beta RB_R)\vec{\d T'}\wedge \vec{\d\phi}+(RB_R-\beta E_\phi)\vec{\d Z'}\wedge\vec{\d\phi}$). But this source is accelerated and thus the axial electric field cannot be transformed away by Lorentz transformation.}

Regarding the invariants \re{inv1} and \re{inv2} it is of interest to see where they vanish. The invariantly defined regions where this occurs are given by the roots of the equations 
\begin{align}
|\vec{E}|^2-|\vec{B}|^2 &=0\,,& \text{Eq.}&\ \re{inv1}\,:& aAxy &= \pm\left( \sqrt{2}\pm 1 \right), &x&=y\,, \\
\vec{E}\cdot\vec{B}&=0\,,&  \text{Eq.}&\ \re{inv2}\,:&  aAxy &= \pm 1\,, &x&=y\,, &y&=0\,.
\end{align}
The roots $x=y$ correspond to asymptotic infinity. These places are indicated in Figs. \ref{fig:accMF0} and \ref{fig:accMF}\subref{3b}; see the captions for details.

\section{Sources} \label{sec:sources}
As mentioned in Introduction, the stationary magic field solves the Maxwell equations with the source in the form of a rigidly rotating relativistic disc. Here we make some remarks on the sources of the accelerated magic field. Assume the field is produced by a 4-current $j^a$ and denote its associated current 3-form by $\vec{J} = j^a\epsilon_{abcd}\, \vec{\d} x^b\wedge\vec{\d} x^c\wedge\vec{\d} x^d$; the Maxwell equations read
\begin{equation}
\vec{\d}\star\vec{F} = 4\pi \vec{J}\,,
\label{eq:MxwF}
\end{equation}
where $\star\vec{F}$ is the 2-form dual to $\vec{F}=F_{ab}\vec{\d}x^a\wedge\vec{\d}x^b$.

Let us multiply Eq. \re{MxwF} from left by a 1-form $\vec{\d} u$ where $u$ is an arbitrary smooth function (e.g., a coordinate) defined on some neighbourhood of a spacetime point and integrate over a 4-volume $\Omega$:
\begin{equation}
\int_{\Omega}\vec{\d} u\wedge\vec{\d}\star\vec{F} = 4\pi\int_{\Omega} \vec{\d} u\wedge\vec{J}\,.
\label{eq:}
\end{equation}
Using Stokes theorem and employing the notation of \CITE{Wald} with $\vec{\epsilon}$ as a 4-volume element, we get
\begin{equation}
-\int_{\p\Omega}\vec{\d} u\wedge\star\vec{F} = 4\pi\int_{\Omega}\vec{\d} u\wedge\vec{J} = -4\pi\int_{\Omega} \vec{\d} u(\vec{j}) \, \vec{\epsilon}\,.
\label{eq:eqS}
\end{equation}

The source we are interested in is on a rotating disc given in ``accelerated'' oblate spheroidal coordinates (metric \re{accOSph}) by $t=\,$const, $y=-\infty$, $x\in ( -1,\,1)$ and $\phi\in\langle 0,\,2\pi )$. Thus the 4-volume of integration will be a region infinitesimally thin around the point of the disc, i.e., $r\in(0,\,r_\epsilon)$ (see transformation \re{}).

The flow of the field \re{eqS} can be evaluated directly in the limit $y\rightarrow -\infty$. At the l.h.s. of \re{eqS} we get
$$-\int_{\p\Omega}\left( \vec{\d} u\wedge\star\vec{F} \right) = -\int_{\p\Omega}\left( \vec{\d} u\wedge\star\vec{F} \right)_{tx\phi} \d t\,\d x\,\d\phi\,.$$
The discussion of the r.h.s. of \re{eqS} is more complicated. The sources are represented by distributions located only at $y=-\infty$. In the limit we have to divide $\int_{\infty}^{y_\epsilon} u_{,\alpha}j^\alpha \sqrt{-g}\, \d y$ by the volume thickness $\int_{-\infty}^{y_\epsilon} \sqrt{g_{yy}}\, \d y$. We get
$$\lim_{y_{\epsilon}\rightarrow -\infty} \frac{\int_{-\infty}^{y_\epsilon} u_{,\alpha}j^\alpha \sqrt{-g}\, \d y}{\int_{-\infty}^{y_\epsilon} \sqrt{g_{yy}}\, \d y} = \frac{u_{,\alpha}j^\alpha\sqrt{-g}}{\sqrt{g_{yy}}}\,.$$
(The same result follows directly from calculus of distributions -- assuming the charge to be located at some 3-surface, we can use the identity
$\int_V f(\vec{r})\, \delta( h(\vec{r}))\,\d^n \vec{r} = \int_{\p V} \frac{f(\vec{r})}{|\nabla h|}\,\d^{n-1}\vec{r}$.)

Collecting the results we obtain the relation 
\begin{equation}
-\int_{\p\Omega}\left( \vec{\d} u\wedge\star\vec{F} \right)_{tx\phi} \d t\,\d x\,\d\phi = 4\pi \int \vec{\d} u(\vec{j})\, \frac{a^2x}{A\left( 1+a^2A^2 \right)}\,\d x\,\d t\,\d\phi
\label{eq:}
\end{equation}
from which
\begin{equation}
4\pi j^\alpha\,u_{,\alpha} = \left( \vec{\d} u\wedge\star\vec{F} \right)_{tx\phi} \frac{A\left( 1+a^2A^2 \right)}{a^2x}\,.
\label{eq:source}
\end{equation}
If we now choose function $u$ to be coordinate $t$ we get the $j^t$ component of the surface 4-current and, analogically, with $u=\phi$ we find the $j^\phi$ component. As a result we receive the 4-current in the form
\begin{equation}
\vec{j} = \frac{1}{2\pi}\frac{q}{a^2x^3} \left(A\vec{\p_t} -\frac{1}{a}\, \vec{\p_{\phi}} \right).\label{eq:4current}
\end{equation}
This is the generalization of the surface 4-current on the disc producing the stationary magic field (cf. \re{mf4c}). On the other disc, accelerating in the opposite direction, we find the opposite signs. As in the stationary case, there are also two singular rings at the rims of the discs with charges and currents of the opposite signs which rectify the surface charges and currents on the discs. 

Transforming now from accelerated oblate spheroidal coordinates via Rindler coordinates to Lorentzian coordinates we subsequently obtain
\begin{multline}
\vec{j} = j^{t}(x) \vec{\p_t} + j^{\phi}(x) \vec{\p_{\phi}} = j^{t}(\zeta,\rho) \vec{\p_t} + j^{\phi}(\zeta,\rho) \vec{\p_{\phi}}\\ 
= Zj^t\!\left( Z^2-T^2,\rho \right)\vec{\p_T}+Tj^t\!\left( Z^2-T^2,\rho \right)\vec{\p_Z}+j^{\phi}\!\left( Z^2-T^2,\rho \right)\vec{\p_{\phi}} \\
 = Zj^t\left[ \vec{\p_T} +\frac{T}{Z}\,\vec{\p_Z} +\frac{j^\phi}{j^t}\frac{\rho}{Z}\,\left( \frac{1}{\rho}\vec{\p_\phi} \right) \right] \\
= Z j^t \left[ \vec{\p_T}+\frac{T}{\sqrt{\zeta^2+T^2}}\,\vec{\p_Z}-\frac{1}{aA}\frac{\rho}{\sqrt{\zeta^2+T^2}}\left( \frac{1}{\rho}\,\vec{\p_\phi} \right) \right].
\end{multline}
Here the 4-current is written in the form $j^\mu=\sigma\left(1,\vec{v} \right)$ in the physical (orthonormal) basis. The velocity is $\vec{v} = (T/Z)\,\vec{e_Z}-(1/( aA ))\,(\rho/Z)\, \vec{e_\phi}$. Its magnitude is given by 
\begin{equation}
|\vec{v}| = \frac{1}{a}\sqrt{\frac{\rho^2+(aAT)^2}{1-A^2\rho^2+A^2T^2}}\,,
\label{eq:3v}
\end{equation}
which yields $|\vec{v}|=AT/\sqrt{1+A^2T^2}$ at the centre of the disc, $\rho=0$, whereas at the rim, i.e. for $\rho_{\textrm{rim}}=a/\Gamma$, we get $|\vec{v}|=1$ for all times $T$. Hence, the charges at the outer rim move with the speed of light as in the case of an unaccelerated relativistic disc producing the stationary electromagnetic magic field. The ``unaccelerated'' case immediately follows from \re{3v} with $A=0$. Then $|\vec{v}|=v_\phi=\rho/a$ which at the rim $\rho_{\textrm{rim}}=a$ gives $|\vec{v}|=1$.
 
The trajectories of individual particles in the disc (for which $\rho=\,$const and $\zeta=\,$const) are easy to obtain by solving the system of ODEs given by $\d \vec{r}(T) / \d T = \vec{v}$. The solution is
\begin{eqnarray}
Z &=& \sqrt{\zeta^2+T^2}\,, \\
\phi &=& -\frac{1}{aA}\,\ln\left( T+\sqrt{\zeta^2+T^2} \right),
\label{}
\end{eqnarray}
where values of $\rho$ and $\zeta$ for particles on the disc are given by Eqs. \re{discZ}\,--\,\re{discR}. In those equations the coordinate $x$ parameterizes the ``distance from the centre of the disc''. The spiral character of the motion of two individual particles of the discs located symetrically with respect to the equatorial plane and accelerated in opposite directions is illustrated in Fig. \ref{fig:accMF}\subref{spiral}.

The angular velocity of rotation of the disc as seen at spacelike surfaces $T=\,$const is 
\begin{equation}
\omega = \frac{\d\phi}{\d T} = -\frac{1}{aA}\frac{1}{Z} = -\frac{1}{a}\frac{1}{\sqrt{1+A^2(T^2-\rho^2)}}\,.
\label{eq:}
\end{equation}
Thus the disc asymptotically stops its rotation in a given inertial frame, as it moves almost with the speed of light in the direction of the $Z$-axis at large $T$.

The bending of the discs appears to be plausible on physical grounds. The discs rotate but their shape remains rigid in Rindler's coordinates, or, equivalently, in any inertial frame in which the centre of the disc is momentarily at rest. (The field and the source are invariant with respect to the boosts along the $Z$-axis.) The outer parts of each of the discs move with higher velocities transversal to the linear acceleration along the $Z$-axis than the inner parts. To preserve the same shape in any of the frames in which the centre is at rest, the outer parts must experience a higher acceleration along the $Z$-axis than the inner parts. This is the case when the discs are bent.

\begin{figure}
\begin{center}
\includegraphics[keepaspectratio, width=0.8\textwidth]{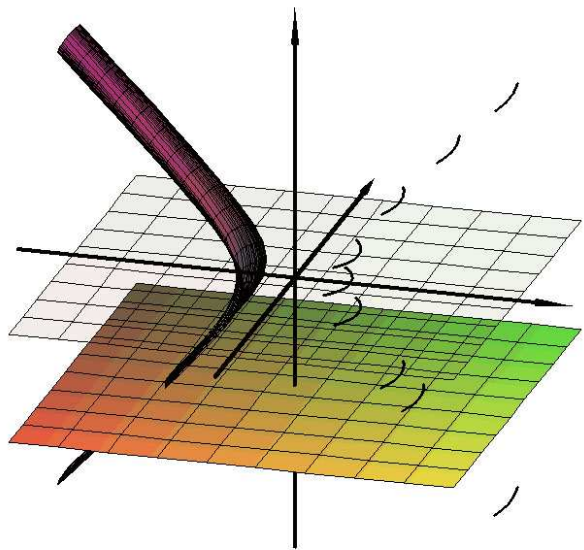}
\end{center}
\caption{The spacetime diagram of the world tube (left) and the shape of discs uniformly accelerated in opposite directions producing the accelerating electromagnetic magic field.}
\label{fig:3d}
\end{figure}

In Fig. \ref{fig:3d} we illustrate the worldtube of the disc moving along $Z<0$ and a number of snapshots of the disc moving along $Z>0$. The bending of the disc for small $T$ and the flattening of the disc for $T\rightarrow \pm \infty$ is here demonstrated manifestly.

\begin{footnotesize}
\bf{Acknowledgements:}
J.B. remains grateful to J\"{u}rgen Ehlers for his support of the interactions with his group in Garching and, later, at the Albert Einstein Institute in Golm, for many (not only scientific) discussions and for generous friendship for more than 25 years. 

We would not know about intriguing properties of the magic field without many inspiring contacts with Donald Lynden-Bell.

Our work is supported by the grants MSM0021620860 of the Ministry of Education of the Czech Republic and by the grant No. LC06014 of the Centre of Theoretical Astrophysics. J.B. also acknowledges the partial support from the Grant GA\v{C}R 202/09/0772.
\end{footnotesize}

\appendix

\section{Zero acceleration limit of the rotating charged C-metric} \label{app:limA}
The rotating charged C-metric we begin with reads
\begin{multline}
\d s^2 = \frac{1}{A^2(x-y)^2}\,\biggl\{ \frac{\mathcal{G}(y)}{1+\left( aAxy \right)^2}\;\Bigl[ \left( 1+a^2A^2x^2 \right)K\d t+aA\left( 1-x^2 \right)K\d\phi \Bigr]^2\\
-\frac{1+\left( aAxy \right)^2}{\mathcal{G}(y)}\,\d y^2 +\frac{1+\left( aAxy \right)^2}{\mathcal{G}(x)}\,\d x^2 \\
+\frac{\mathcal{G}(x)}{1+\left( aAxy \right)^2}\;\Bigl[ \left( 1+a^2A^2y^2 \right)K\d\phi+aA\left(y^2-1\right)K\d t \Bigr]^2\biggr\}
\label{eq:ArCMr}
\end{multline}
and the 4-potential, given by Eq. \re{rA}, is
\begin{equation}
\vec{\fp} = \frac{Kqy}{1+\left( aAxy \right)^2}\,\left[ \left(1+a^2A^2x^2\right)\vec{\d} t+aA\left( 1-x^2 \right)\vec{\d}\phi \right]\,.
\label{eq:ArA}
\end{equation}

Applying transformation (cf. also \CITE{HT2})
\begin{equation}
t\rightarrow tA\,,\qquad x\rightarrow \cos\theta\,,\qquad y\rightarrow -\frac{1}{Ar}\,,
\label{eq:trfAa}
\end{equation}
so that $\d t \rightarrow A\,\d t$, $\d x \rightarrow -\sin\theta\,\d\theta$, $\d y \rightarrow \frac{1}{Ar^2}\,\d r$, the relevant functions in the metric become
\begin{eqnarray}
\mathcal{G}(x) &\ \rightarrow\ & \sin^2\theta \left( 1+r_{\ssp}A\cos\theta \right)\left( 1+r_{\ssm}A\cos\theta \right), \\
\mathcal{G}(y) &\ \rightarrow\ & \frac{A^2r^2-1}{A^2r^4} \left( r-r_{\ssp} \right)\left( r-r_{\ssm} \right) = \frac{1}{A^2r^4} \left( A^2r^2-1 \right) \Delta\,,\\
A^2(x-y)^2 &\ =\ & \frac{1}{r^2}\left( Ar\cos\theta+1 \right),\\
1+\left(aAxy\right)^2 &\ =\ & \frac{1}{r^2} \Sigma\,,
\label{}
\end{eqnarray}
where $\Delta=(r-r_{\ssp})(r-r_{\ssm})=r^2-2Gm+a^2+Gq^2$ and $\Sigma = r^2+a^2\cos^2\theta$ are standard notations (see, e.g., \CITE{MTW}, \CITE{Wald}).

Substituting the above results into the metric and taking the limit $A\rightarrow 0$ (we see from Eq. \re{rDefAo} that $K\rightarrow 1$ in physically relevant cases), we get
\begin{multline}
\d s^2 = r^2\biggl\{ -\frac{\Delta}{r^2\Sigma}\left[ \d t+ a\sin^2\theta \d\phi \right]^2 +\frac{\Sigma}{r^2\Delta}\,\d r^2+\frac{\Sigma}{r^2}\,\d\theta^2 \\
+ \frac{r^2\sin^2\theta}{\Sigma}\left[ \left( 1+\frac{a^2}{r^2} \right)\d\phi+\frac{a}{r^2}\,\d t^2 \right]^2 \biggr\}\,.
\end{multline}
After simple rearrangements this yields
\begin{multline}
\d s^2 = \frac{-\Delta+a^2\sin^2\theta}{\Sigma}\,\d t^2 +\frac{\Sigma}{\Delta}\,\d r^2 +\Sigma\,\d\theta^2 + 2a\,\frac{\left( r^2+a^2\right)-\Delta}{\Sigma}\,\sin^2\theta\,\d t\d\phi \\
+ \frac{\left( r^2+a^2 \right)^2-a^2\Delta\sin^2\theta}{\Sigma}\,\sin^2\theta\,\d\phi^2\,,
\label{eq:}
\end{multline}
which is the standard form of the Kerr-Newman metric in Boyer-Lindquist coordinates (see, e.g., \CITE{MTW}, \CITE{Wald}). After using the transformation \re{trfAa}, the electromagnetic field \re{ArA} becomes
\begin{equation}
\vec{\fp} = -\frac{qr}{\Sigma}\left( \vec{\d} t+a\sin^2\theta\,\vec{\d}\phi \right).
\label{eq:}
\end{equation}

\newpage
\textbf{\emph{Electromagnetic magic field}}
\vspace{.5em}

As discussed in the main text, the magic field is the special-relativistic limit of the Kerr-Newman solution. The metric in this limit goes over over to the flat spacetime metric in oblate spheroidal coordinates\footnote{The coordinates $r,\,\theta,\,\phi$ are, in fact, the ``quasi-spherical spheroidal coordinates''. They are connected with the standard dimensionless spheroidal coordinates $\mu,\,\nu$ by relations $r=a\sinh\mu$ and $\cos\theta=\sin\nu$ (see, e.g., \CITE{DLB-Magic}, \CITE{DLB-ElM-PrD} for more details).}:
\begin{equation}
\d s^2 = -\d t^2 + \frac{r^2+a^2\cos^2\theta}{r^2+a^2}\,\d r +\left( r^2+a^2\cos^2\theta \right)\d\theta^2 +\left( r^2+a^2 \right)\sin^2\theta\,\d\phi^2 \,.
\label{eq:}
\end{equation}
The magic field is given by the 4-potential 
\begin{equation}
\vec{\fp} =  \frac{qr}{r^2+a^2\cos^2\theta}\left( -\vec{\d} t +a\sin^2\theta\vec{\d}\phi \right).
\label{eq:}
\end{equation}
The electromagnetic field reads
\begin{multline}
\vec{F} =  \frac{q}{\left( r^2+a^2\cos^2\theta \right)^2} \Bigg\{\left( r^2-a^2\cos^2\theta \right)\vec{\d}r\wedge\left( \vec{\d}t-a\sin^2\theta\,\vec{\d}\phi \right) \\
+2ar\cos\theta\sin\theta\,\vec{\d}\theta\wedge \left[ \left( r^2+a^2 \right)\vec{\d}\phi-a\,\vec{\d}t\right] \Bigg\}.
\label{eq:}
\end{multline}
The 4-current on the surface of the disc (given by $r=0$, i.e., $a^2-r_{\mathrm{spherical}}^2=a^2\cos^2\theta$ in spheroidal coordinates) is given by 
\begin{equation}
\vec{j} = \frac{1}{2\pi}\frac{q}{a^2\cos^3\theta}\left( -\vec{\p_t}+\frac{1}{a}\,\vec{\p_\phi} \right),
\label{eq:mf4c}
\end{equation}
see, e.g., \CITE{DLB-Magic}, \CITE{DLB-ElM-PrD}.

\section{Born's solution} \label{app:Born}
We demonstrate how to obtain Born's solution (see, e.g., \CITE{R}) for two uniformly accelerated point monopole charges (cf. Fig. \ref{fig:1}) as a flat spacetime limit of non-rotating charged C-metric 
\begin{equation}
\d s^2 = \frac{1}{(x-y)^2}\left[ \frac{\mathcal{G}(y)}{\lambda}\,\d t^2 - \frac{1}{A^2\lambda^2\,\mathcal{G}(y)}\,\d y^2 +\frac{1}{A^2\lambda^2\,\mathcal{G}(x)}\,\d x^2+\frac{\mathcal{G}(x)}{A^2\lambda^2}\,\d \phi^2 \right].
\label{eq:SIcm}
\end{equation}
Here we introduced $t$ (see below) with dimension of time, $[\siun{s}]$, the other coordinates remain dimensionless. Constant $A\,[\siun{m}\cdot\siun{s}^{-2}]$ is the acceleration, $\lambda\,[\siun{m}^{-2}\cdot\siun{s}^2]=c^{-2}$ ($c$ is the speed of light). The structure function reads 
\begin{equation}
\mathcal{G}(\xi) = \left( 1-\xi^2 \right)\left( 1+A\lambda r_{\ssp}\xi \right)\left( 1+A\lambda r_{\ssm}\xi \right),
\label{eq:}
\end{equation}
where
\begin{equation}
r_{\sspm} = G\lambda m + \sqrt{\left(G\lambda m\right)^2 \pm G\lambda^2q^2}\,.
\label{eq:}
\end{equation}
The 4-potential is
\begin{equation}
\vec{\fp} = A\lambda qy \,\vec{\d} t\,.
\label{eq:SIcm4p}
\end{equation}
Above we incorporated the factor $\lambda^{-1/2}A^{-1}$ into the time coordinate $t$ so that it has dimension $[\siun{s}]$, in contrast to the dimensionless time coordinate in the original form of the C-metric \re{rCM}. The field is given by
\begin{equation}
\vec{F} = -A\lambda\,q\,\vec{\d}t\wedge\vec{\d}y\,.
\label{eq:SIf}
\end{equation}
A flat spacetime metric in cylindrical coordinates with standard units, 
\begin{equation}
\d s^2 = -\frac{1}{\lambda}\,\d T^2 + \d Z^2 + \d \rho^2 + \rho^2\d\phi^2\,,
\label{eq:Mink1}
\end{equation}
is obtained in the limit $G\rightarrow 0$ of the C-metric \re{SIcm} by transformation
\rem{\begin{align}
t &= \frac{1}{A\sqrt{\lambda}}\,\atgh\left( \frac{T}{Z\sqrt{\lambda}} \right) & x &= -\frac{\rho^2+Z^2-T^2/\lambda -1/A^2\lambda^2}{\sqrt{\left(\rho^2+Z^2-T^2/\lambda -1/A^2\lambda^2\right)^2+4\rho^2/A^2\lambda^2}} \\
\phi &=\phi & y &= -\frac{\rho^2+Z^2-T^2/\lambda +1/A^2\lambda^2}{\sqrt{\left(\rho^2+Z^2-T^2/\lambda -1/A^2\lambda^2\right)^2+4\rho^2/A^2\lambda^2}}
\label{eq:trfMtoCM}
\end{align}}
\begin{align}
t &= \frac{1}{A\sqrt{\lambda}}\,\atgh\left( \frac{T}{Z\sqrt{\lambda}} \right), & x &= -\frac{u^2}{\sqrt{u^4+4\rho^2(A\lambda)^{-2}}}\,, \\
\phi &=\phi\,, & y &= -\frac{u^2 +2\,(A\lambda)^{-2}}{\sqrt{u^4+4\rho^2(A\lambda)^{-2}}}\,,
\label{eq:trfMtoCM}
\end{align}
where
\begin{eqnarray}
u^2 &=& \rho^2+Z^2-T^2/\lambda -1/A^2\lambda^2\,, \\
\xi_{\ssp} &=& \sqrt{u^4+4\rho^2(A\lambda)^{-2}}\,.
\label{eq:SIuDef}
\end{eqnarray}
The electric and the magnetic fields measured by an observer with 4-velocity $\vec{u}$ are given by
\begin{eqnarray}
E_a &=& F_{ab}u^b\,, \label{eq:Esi}\\
B_a &=& \frac{1}{2}\sqrt{\lambda}\,\epsilon_{abcd}\,u^bF^{cd}\,. \label{eq:Bsi}
\end{eqnarray}
An inertial observer with $\vec{u}=\vec{\p_T}$ measures
\begin{eqnarray}
\vec{E} &=& \frac{q}{A^2\lambda^2}\frac{4}{\xi_{\ssp}^3}\left[ \left( u^2-2\rho^2 \right)\frac{\vec{\p}}{\vec{\p}Z} + 2Z\rho\,\frac{\vec{\p}}{\vec{\p}\rho} \right], \label{eq:BornE} \\
\vec{B} &=& \frac{q}{A^2\lambda^2}\frac{1}{\xi_{\ssp}^3}\,8T\rho^2\,\frac{\vec{\p}}{\vec{\p}\phi}\,. \label{eq:BornB} 
\end{eqnarray}
This is precisely Born's solution for two uniformly accelerated charged monopole particles (see, e.g., \CITE{R}, \CITE{Bi68}).

\bibliography{MF.bib}
\end{document}